\documentclass{aa}  

\usepackage{graphicx}
\usepackage{changes}
\usepackage{txfonts}
\usepackage[flushleft]{threeparttable}
\usepackage{xcolor}
\usepackage[switch]{lineno} %switch for two column
\begin{document} 

%\linenumbers

   \title{LHAASO\,J2108+5157 as a Molecular Cloud Illuminated by a Supernova Remnant}

   %Had the idea on a Thursday, submitted on the Tuesday. Did this paper in less than a week. Should try doing that more often...

   \author{A.M.W. Mitchell
          \inst{1} \thanks{https://orcid.org/0000-0003-3631-5648}
          }

   \institute{Erlangen Centre for Astroparticle Physics,
              Friedrich-Alexander-Universit\"at Erlangen-N\"urnberg, Nikolaus-Fiebiger-Stra{\ss}e 2, 91058 Erlangen, Germany\\
              \email{alison.mw.mitchell@fau.de}
             }

   \date{Received --; accepted --}
 
  \abstract
   {The search for Galactic PeVatrons - astrophysical accelerators of cosmic rays to PeV energies - has entered a new phase in recent years with the discovery of the first Ultra-High-Energy (UHE, $E>100$\,TeV) gamma-ray sources by the HAWC and LHAASO experiments. Establishing whether the emission is leptonic or hadronic in nature, however, requires multiwavelength data and modelling studies. Among the currently known UHE sources, LHAASO\,J2108+5157 is an enigmatic source without clear association to a plausible accelerator, yet spatially coincident with molecular clouds. }
   {We investigate the scenario of a molecular cloud illuminated by cosmic rays accelerated in a nearby supernova remnant (SNR) as an explanation for LHAASO\,J2108+5157. We aim to constrain the required properties of the SNR as well as which of the clouds identified in the vicinity is the most likely association. }
   {We use a model for cosmic ray acceleration in SNRs, their transport through the interstellar medium and subsequent interaction with molecular material, to predict the corresponding gamma-ray emission. The parameter space of SNR properties is explored to find the most plausible parameter combination that can account for the gamma-ray spectrum of LHAASO\,J2108+5157.  }
   {In the case that a SNR is illuminating the cloud, we find that it must be young ($<10$\,kyr) and located within $40-60$\,pc of the cloud. A SN scenario with a low Sedov time is preferred, with a  maximum proton energy of 3\,PeV assumed. No SNRs matching these properties are currently known, although an as yet undetected SNR remains feasible. The galactic CR sea is insufficient to solely account for the observed flux, such that a PeVatron accelerator must be present in the vicinity. }
   {}

   \keywords{Gamma ray: ISM --
                ISM: supernova remnants --
                ISM: clouds --
                (ISM:) cosmic rays
               }

   \maketitle
%\citet vs \citeyear vs \citep
%-------------------------------------------------------------------

\section{Introduction}
\label{sec:intro}

Cosmic Rays (CRs) are energetic particles originating from astrophysical accelerators and continuously arriving at Earth. The all-particle CR spectrum exhibits a spectral softening at $\sim$ 1-3\,PeV, known as the `knee', generally understood to indicate the start of the transition from galactic to extragalactic accelerators being responsible for the bulk of CRs \citep{1964ocr..book.....Ginzburg,2006astro.ph..7109Hillas,2013PhRvD..87h1101A_kascade,2014NuPhS.256..197Parizot}. 
Astrophysical sources capable of accelerating particles to PeV energies are known colloquially as `PeVatrons'. Gamma rays produced as a consequence of particle interactions at the source typically have energies a factor $\sim 1/10$ that of the parent particle population \citep{1970RvMP...42..237Blumenthal}. 
Hence, the detection of gamma rays with $E>100$\,TeV indicates the presence of particles with PeV energies, corresponding to the CR `knee'. 

Definitive evidence for  the presence of PeVatrons in our galaxy has, however, proven elusive. 
Although diffusive shock acceleration of CRs at supernova remnants (SNRs) can account for the energy budget of CRs in our galaxy, their gamma-ray spectra cut-off at energies below 100\,TeV \citep{1978MNRAS.182..147Bell,1983A&A...125..249LagageSNREmax,2013APh....43...56BellReview,2018A&A...612A...3HESS_snrpop}. 
This indicates that active acceleration of particles to PeV energies is not occurring at these SNRs, although the detection of the characteristic `pion-bump' signature of neutral pion decay in several SNRs indicates that the emission is hadronic in origin \citep{2013MNRAS.431..415BellSNR,2013Sci...339..807A_FermiPion,2016ApJ...816..100Jogler}. 

Indications for PeVatron activity from the galactic centre were found \citep{2016Natur.531..476HESSGC}, yet only in recent years have experimental facilities been capable of measuring gamma-rays with energies $>100$\,TeV. Water Cherenkov and particle detector based facilities in particular, such as HAWC \citep{2017ApJ...843...39A_hawcCrab}, LHAASO \citep{Aharonian_2021_lhaasoCrab} and Tibet-AS$\gamma$ \citep{2009ApJ...692...61A_tibetAScrab} have contributed significantly to this advance. 

Until 2023, the ultra-high-energy (UHE, $>100$\,TeV) gamma-ray sky comprised a mere $\sim15$ sources, with the Crab nebula one of the first identified \citep{2019PhRvL.123e1101A_tibetCrab,2021Sci...373..425LhaasoCrab}. A further 31 UHE sources were announced in the first LHAASO catalogue \citep{2023arXiv230517030Cao_1lhaaso}. %43 total, less 12 previously announced. 
Twelve UHE sources were reported by LHAASO in 2021 \citep{2021Natur.594...33CaoUHE}, the majority of which are spatially coincident with known Very-High-Energy (VHE, $\gtrsim 100$\,TeV) gamma-ray sources. In particular, several associations with energetic pulsar wind nebulae from which the emission is understood to be dominantly leptonic in origin. 
Despite Klein-Nishina suppression of inverse Compton scattering at the highest energies, this suppression is relaxed in the case of high radiation field environments, and a leptonic scenario remains a viable interpretation for the UHE sources associated with known energetic pulsars \citep{2009A&A...497...17Vannoni,2021ApJ...908L..49BreuhausUHE,2022A&A...660A...8Breuhaus}. 

There is, however, one source reported in \cite{2021Natur.594...33CaoUHE}, for which the gamma-ray emission was first discovered at UHE and without any known counterpart accelerators, such as pulsars or supernova remnants (SNRs). 
LHAASO\,J2108+5157 is an enigmatic source, spatially coincident with molecular clouds yet with the accelerator mechanism remaining unidentified \citep{2021ApJ...919L..22Cao}. 
In the wake of the LHAASO discovery, follow-up observations were conducted by several facilities, including in the radio and X-ray bands as well as by gamma-ray experiments. The Fermi-LAT source 4FGL\,J2108.0+5155 is spatially coincident with the UHE emission, but due to the differing spectral properties a physical association remains unclear \citep{2020ApJS..247...33A_4fgl}. 
A re-analysis of the Fermi-LAT data found a potential spatial extension of $0.48^\circ$ angular size of the source, designated 4FGL\,J2108.0+5155e \citep{2021ApJ...919L..22Cao}. 

A $3.7\sigma$ signal of gamma-ray emission was measured at $E>3$\,TeV by the Large Sized Telescope (LST-1), a prototype telescope for the forthcoming Cherenkov Telescope Array (CTA) 
 \citep{2013APh....43....3A_CTAintro}. They derive upper limits in the energy range 0.32\,TeV to 100\,TeV that considerably constrain model scenarios for the origin of the emission \citep{2023A&A...673A..75A_MWL_LST}.  

The HAWC observatory recently reported a $\sim7\,\sigma$ detection in $\sim2400$ days of data \citep{HAWCj2108_icrc}. However, observations and analysis by the VERITAS IACT array did not result in a detection, with constraining upper limits being reported, consistent with those from the LST-1 \citep{HAWCj2108_icrc,2023A&A...673A..75A_MWL_LST}. 

Although there is little observational evidence for SNRs currently acting as PeVatrons, it remains feasible that SNRs act as PeVatrons only for a comparatively short period during their lifetimes, such that the rate of currently detectable SNR PeVatrons is low \citep{2020APh...12302492Cristofari_rate}. 
Particle escape from the shock region occurs in an energy-dependent manner, such that the most energetic particles will also be the first to leave the shock region \citep{2019MNRAS.490.4317Celli}. 
Evidence for PeV particles may therefore be found not at the location of the accelerator, but rather from subsequent interactions of these particles with target material in the ambient medium, such as nearby molecular clouds \citep{GabiciAharonianPeVatron07,2009MNRAS.392..240MorlinoRXJ,Inoue2012ApJ...744...71I,2019MNRAS.487.3199CelliClumpy}. 
This scenario has been proposed as a possible explanation for the UHE emission from LHAASO\,J2108+5157 \citep{2021ApJ...919L..22Cao,2022ApJ...926..110Kar,2023MNRAS.521L...5DeSarkar,2023A&A...673A..75A_MWL_LST}. 

In contrast to previous models for LHAASO\,J2108+5157, in this study we scan the parameter space in two free variables, namely SNR age and the distance between the cloud and the SNR, to determine the range of plausible values for the required properties of the responsible SNR. We investigate the influence of uncertainties in the cloud properties on the resulting gamma-ray flux for the best-matched models. The corresponding expected neutrino flux is estimated, and the plausibility of the best-matched models is discussed. 

%--------------------------------------------------------------------

\section{Method}
\label{sec:method}

We adopt the model of \cite{2021MNRAS.503.3522Mitchell,2023MNRAS.520..300Mitchell_err}, based on \cite{AA96} and \cite{Kelner06}, to investigate the scenario of a SNR illuminating molecular clouds as a possible explanation for LHAASO\,J2108+5157. Whilst there are several clouds identified in the vicinity, we focus on clouds that are spatially coincident with the $\gamma$-ray emission and located closest to the best-fit centroid of LHAASO\,J2108+5157 at ($l=92.2148^\circ$,$b=2.9359^\circ$). Cloud 4607 from the \cite{MDsurvey17} catalogue based on data from the $^{12}$CO survey of \citet{Dame01} has been considered in previous models of the region \citep{2022ApJ...926..110Kar,2023MNRAS.521L...5DeSarkar}, whilst recently a newly identified cloud has been detected in the region \citep{2023PASJ...75..546DeLaFuente}. Adopting the convention of prior works, we henceforth refer to these two clouds as MML[2017]4607 and FKT[2022] respectively. 
Table \ref{tab:clouds} summarises the key physical properties of the clouds relevant for this study. 

%--------------------------------------------------- One column table
   \begin{table}
      \caption[]{Properties of the molecular clouds considered in this study. $d$ is the distance to the cloud and $n$ is the total number density of hydrogen gas. }
         \label{tab:clouds}
         \begin{tabular}{l|cc}
        Cloud  & MML[2017]4607 & FKT[2022] \\
            \hline
        ($l$ (deg), $b$ (deg)) & (92.272, 2.775) & (92.4, 3.2) \\
        $d$ (kpc) & 3.28 & 1.7 $\pm$ 0.6 \\
        $n$ (cm$^{-3}$) & 30 & 37 $\pm$ 14 \\
        diameter (deg) & 0.5 & 1.1 $\pm$ 0.2 \\
        \hline
         \end{tabular}
   \end{table}

For convenience, we summarise here the key features of the model from \cite{2021MNRAS.503.3522Mitchell,2023MNRAS.520..300Mitchell_err} adopted for this work. 
\begin{itemize}
    \item[--] Protons are accelerated impulsively with a power-law spectrum of slope $\alpha$.
    \item[--] The particle probability density function $f(E,r',t')$ is taken from equation (3) of \cite{AA96}, and is a function of the particle energy $E$, distance travelled from the SNR $r'$ and time since escape from the SNR $t'$.
    \item[--] The SNR radius, $R_{\rm SNR}$, expands with time (t) adiabatically during the Sedov-Taylor phase as $R_{\rm SNR}\propto t^{2/5}$ \citep{1999ApJS..120..299Truelove}.
    \item[--] Particles escape from the SNR at a time $t_{\rm esc}$ in a momentum-dependent manner, following $t_{\rm esc}\propto (p/p_M)^{1/\beta}$ where $p_M$ is the maximum particle energy reached, assumed to be 3\,PeV/c at the Sedov time, $t_{\rm sed}$ \citep{2019MNRAS.490.4317Celli}. 
    \item[--] Particles are either transported diffusively through the ISM to reach the cloud or are injected directly into the cloud if the SNR is sufficiently expanded. 
    \item[--] Diffusion within the intervening ISM is assumed to be slow with respect to the Galactic average value due to the local accelerator activity \citep{Gabici07}.
    \item[--] Within the cloud, diffusion is suppressed with respect to the ISM by a factor $\chi$ that relates to local turbulence. %local to the cloud
\end{itemize}
For our default scenario, the Sedov time is assumed to commence at $1.6$\,kyr corresponding to the case of type II supernovae. In the case of type IA supernovae, the Sedov time commences at a mere 234\,yr, which is considered as an alternative scenario \citep{2019MNRAS.490.4317Celli}.

Details of the SNR forward shock interaction with the cloud, in the case that the SNR is in close proximity and sufficiently evolved, are neglected. 
The diffusion coefficient $D(E)$ is considered to have a power-law dependence on the energy $E$ as:

\begin{equation}
D(E) = \chi D_0 \left(\frac{E /\mathrm{GeV} }{B(n)/3\mathrm{\mu G}}\right)^{\delta}~,
\label{eq:diffcoeff}
\end{equation}

\noindent where $\delta$ and $\chi$ relate to the properties of the magnetic field turbulence in the region, and $B(n)$ describes the dependence of the magnetic field strength $B$ on the cloud density $n$ (see \cite{2021MNRAS.503.3522Mitchell}).  Values adopted for $\delta$, $\chi$ and the diffusion coefficient normalisation $D_0$ at 1\,GeV are listed in table \ref{tab:modelpars}. Within the ISM, $\chi$ is taken to be 1, whilst a value of 0.1 is adopted to account for suppressed diffusion within the clouds. 
From the above ingredients, the particle spectrum as a function of energy $E$, age $t$ and distance from the accelerator $r$ is obtained, $f(E,r',t')$.

\begin{table}[]
    \centering
    \caption{Assumed parameters of the model. The value adopted for $D_0$ is taken from \cite{Gabici07}. }
    \begin{tabular}{lcl}
    Description & Parameter  & Value \\
    \hline
    Proton power law spectral index & $\alpha$    & 2.0 \\
    Index characterising energy & $\delta$ & 0.5 \\
    -dependence of diffusion & & \\
    Index characterising momentum & $\beta$ & 2.5 \\
    -dependence of particle escape & & \\
    Diffusion suppression factor & $\chi$ & 0.1 or 1 \\
    due to turbulence & & \\
    Diffusion coefficient & $D_0 $ & $3\times10^{26}\frac{\rm cm^2}{\rm s}$ \\
    normalisation at 1\,GeV & & \\
    ISM particle density & $n_{\rm ISM}$ & 1\,cm$^{-3}$ \\
    Maximum particle momentum & $p_M$ & 3\,PeV/c \\
    Sedov time (type II SNR) & $t_{\rm sed}$ & 1.6\,kyr \\
    Sedov time (type Ia SNR) & $t_{\rm sed}$ & 234\,yr \\
    \end{tabular}
    \label{tab:modelpars}
\end{table}

Experimental measurements are, however, bound to neutral messengers such as $\gamma$-rays and neutrinos as the signatures for the presence of energetic hadronic particles. For comparison to data, the proton spectrum can then be converted into a gamma-ray emissivity $\Phi_\gamma(E_\gamma,r',t')$ (in ph~cm$^{-3}$~s$^{-1}$~TeV$^{-1}$) by using the expressions from \cite{Kelner06}:

\begin{equation}
\Phi_\gamma (E_\gamma,r',t') = cn \int_{E_\gamma}^{\infty} \sigma_{\mathrm{inel}}(E)f(E,r',t')F_\gamma \left(\frac{E_\gamma}{E},E\right)\frac{dE}{E}~,
\label{eq:gflux}
\end{equation}%n_H
for which we adopt the parameterisation of the inelastic cross-section for proton-proton interactions $\sigma_{\mathrm{inel}}(E)$ from \cite{Kafexhiu14}, noting that due to high uncertainties below $\sim\,100$\,GeV, we take this as an energy threshold and restrict our model predictions to energies $>100$\,GeV only. 

Lastly, we obtain the $\gamma$-ray flux $F(E_\gamma,t)$ at a distance $d$ away from the cloud (i.e. at Earth) taking into account the volume of the molecular cloud traversed by particles $V_c$ via:

\begin{equation}
    F(E_\gamma,t) = \Phi_\gamma (E_\gamma,t) V_c / (4\pi d^2)~.
    \label{eq:cloudflux}
\end{equation}

The diffusive galactic CR flux permeates the entire Galaxy, and as such will also contribute to the total particle flux interacting with the molecular clouds. To take this contribution into account, we include the proton flux as measured by the Alpha Magnetic Spectrometer on the International Space Station \citep{PhysRevLett.114.171103_AMS}. This flux is added to the particle flux arriving at the cloud, $f$ in equation \eqref{eq:gflux}, enabling the relative contributions of accelerator and the diffuse CR sea to be evaluated. 

\cite{Kelner06} also provide expressions for the neutrino production via charged pion and muon decay via $F_\nu\left(\frac{E_\nu}{E},E\right)$ for the total production of electron and muon neutrinos from the same proton interactions. By analogy with equations \eqref{eq:gflux} and \eqref{eq:cloudflux} the corresponding total neutrino flux can be obtained.

In the next section, we use this model to generate predictions for the gamma ray flux arising from a hypothetical SNR illuminating the molecular clouds identified in the vicinity of the LHAASO\,J2108+5157. 
Additionally, we consider the contribution from the galactic CR sea, to establish whether it is sufficient to account for the observed gamma-ray emission without requiring a nearby accelerator.
The model is compared to measurements from LHAASO and HAWC, and upper limits from the LST-1 and VERITAS \citep{2021ApJ...919L..22Cao,2023A&A...673A..75A_MWL_LST,HAWCj2108_icrc}. 

   \begin{figure*}
   \centering
   \includegraphics[width=\columnwidth]{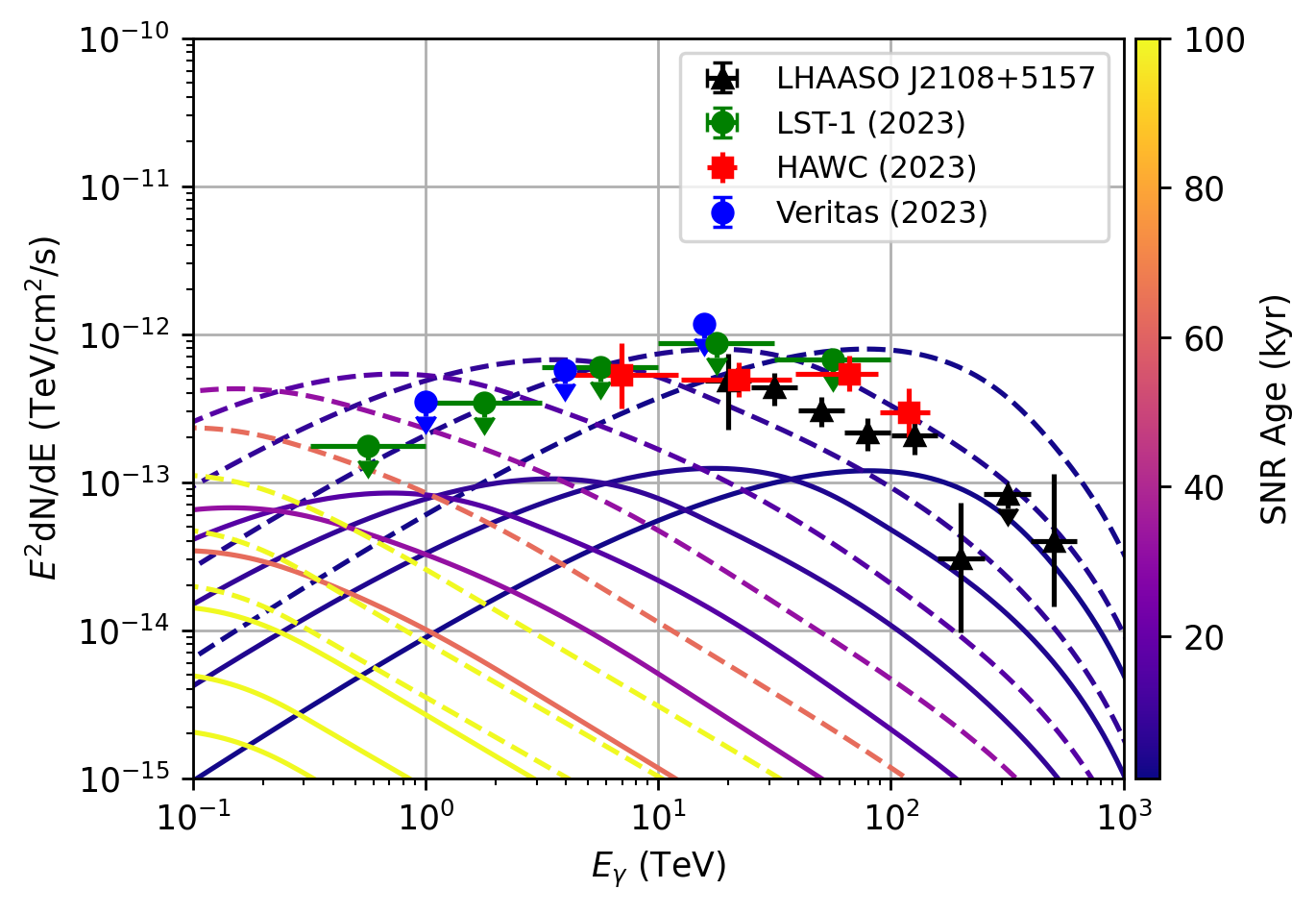}
   \includegraphics[width=\columnwidth]{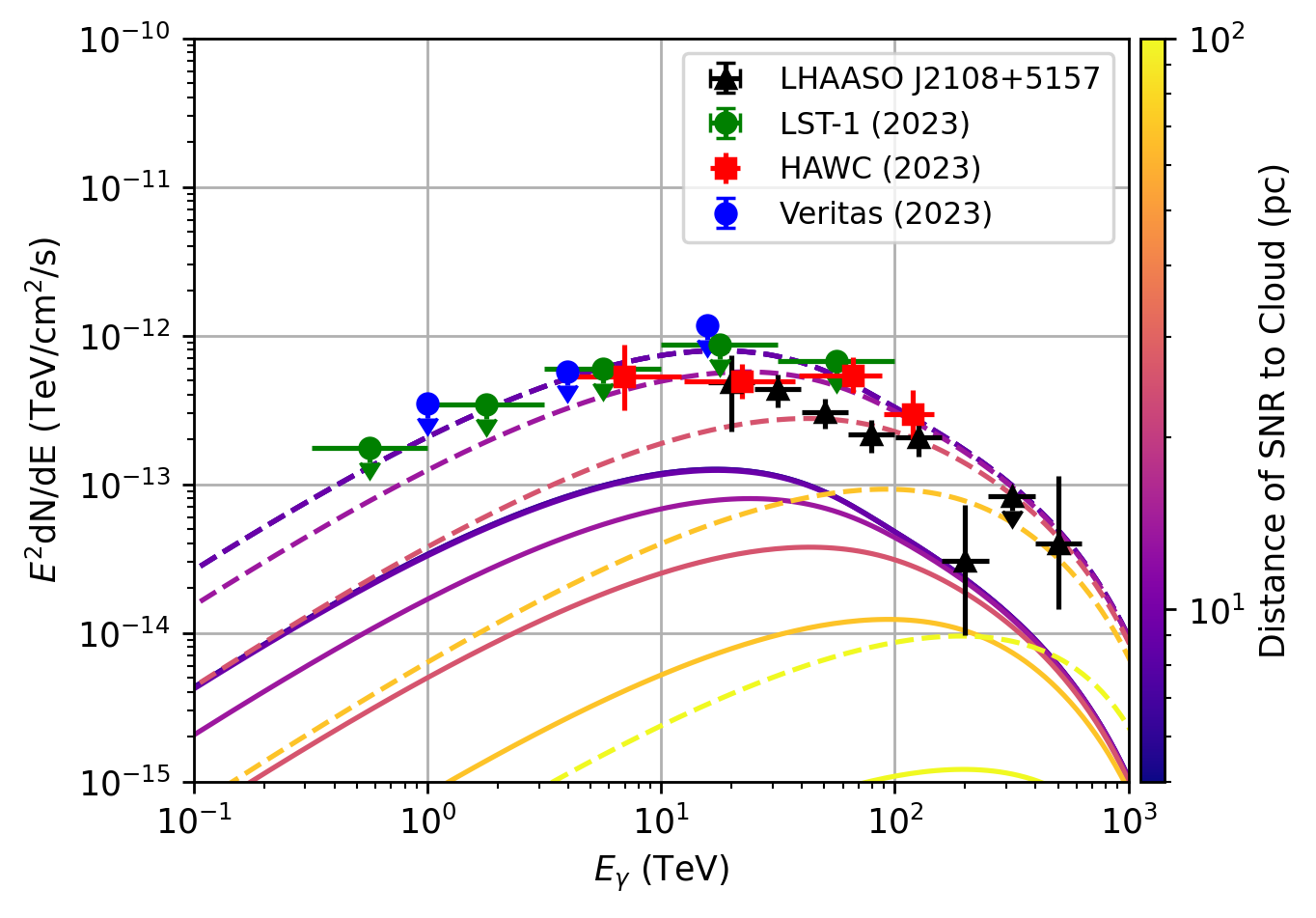}
   \caption{Dependence of gamma-ray flux on properties of a type II supernova. Left: Variation in supernova age for a fixed distance of 24\,pc. Right: Variation with distance between the SNR and the cloud for a fixed age of 4\,kyr. Solid lines correspond to MML[2017]4607 and dashed lines correspond to FKT[2022]. }
              \label{fig:typeII}%
    \end{figure*}

   \begin{figure*}
   \centering
   \includegraphics[width=\columnwidth]{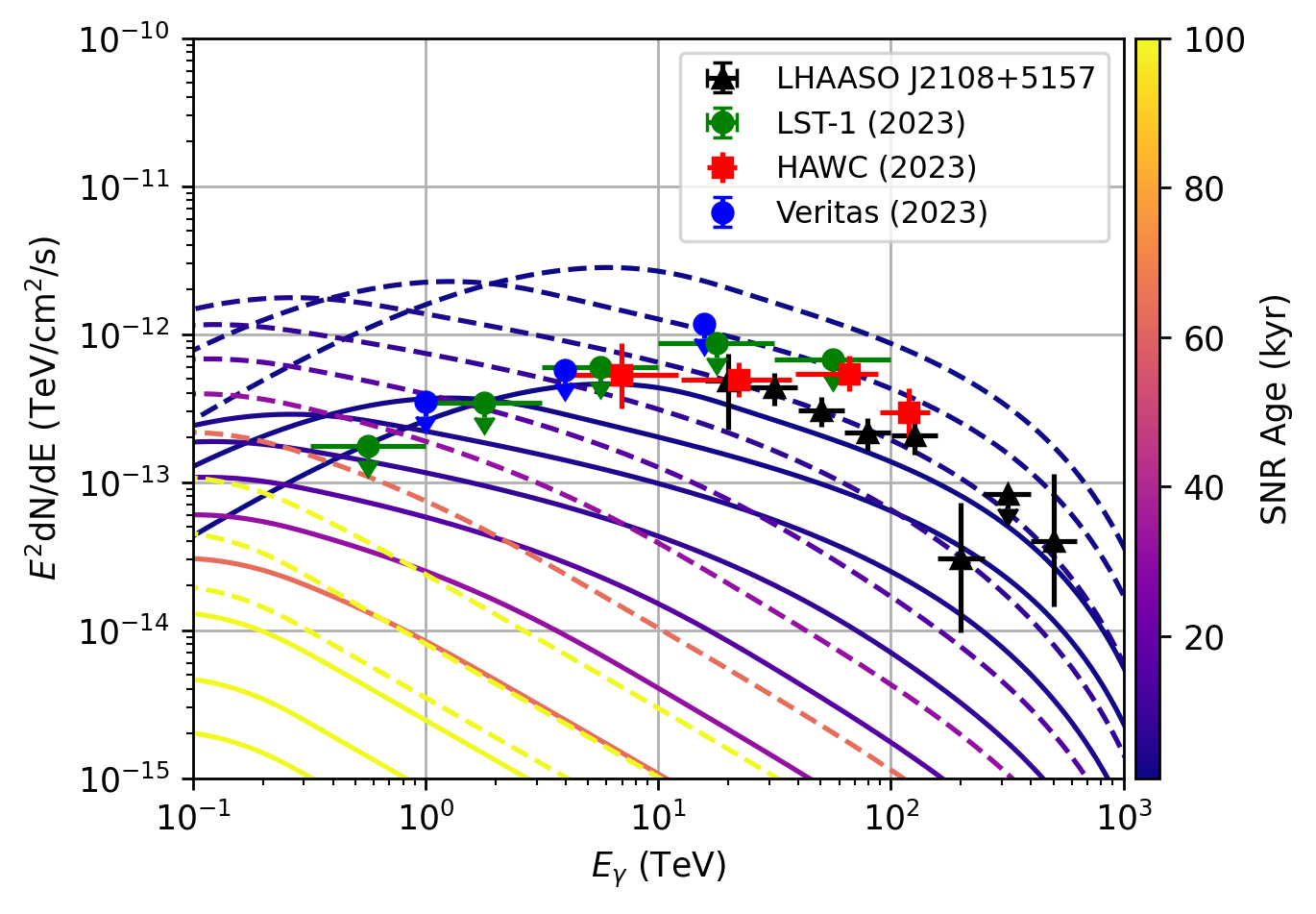}
   \includegraphics[width=\columnwidth]{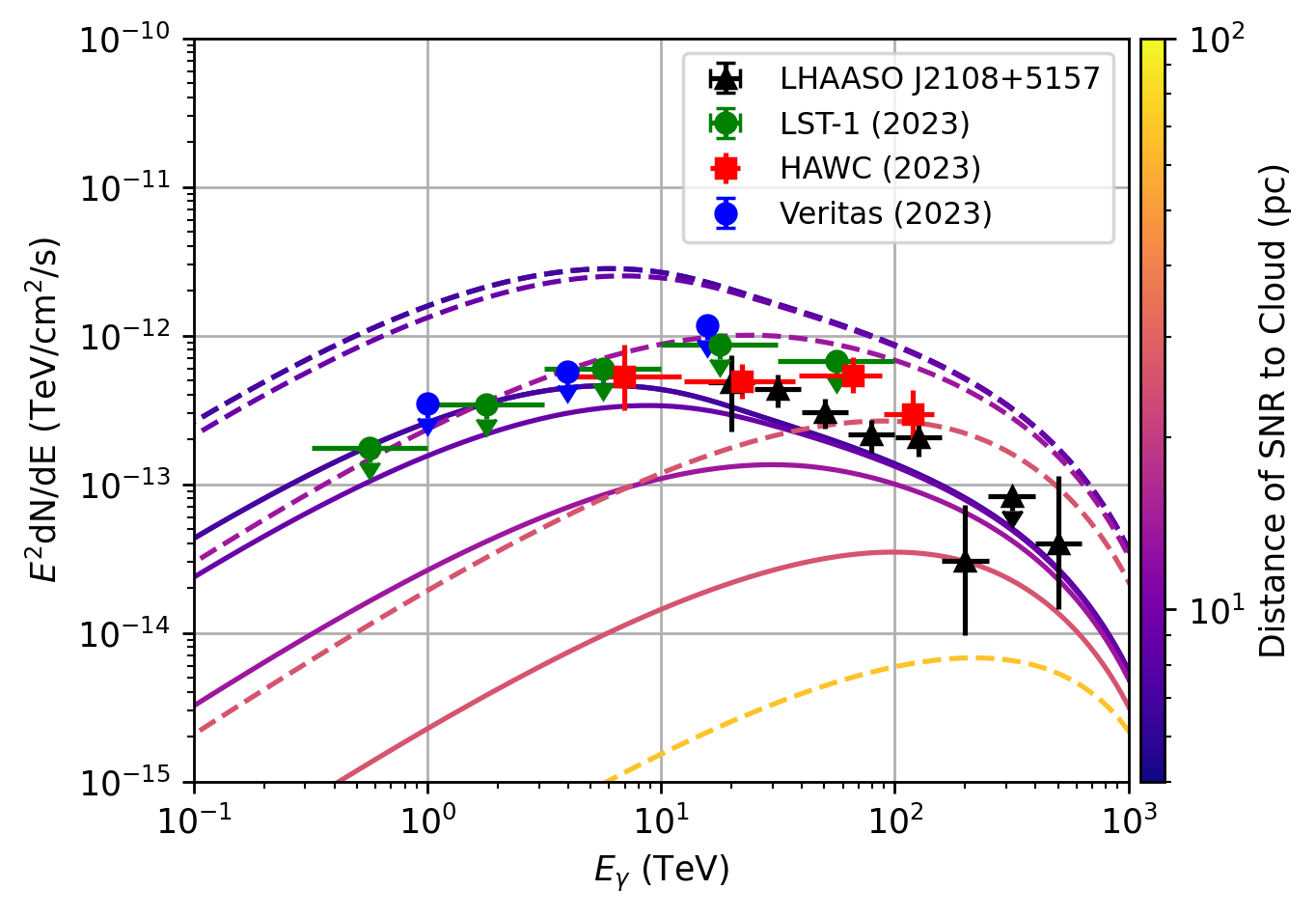}
   \caption{Dependence of gamma-ray flux on properties of a type Ia supernova. Left: Variation in supernova age for a fixed distance of 10\,pc. Right: Variation with distance between the SNR and the cloud for a fixed age of 1\,kyr. Solid lines correspond to MML[2017]4607 and dashed lines correspond to FKT[2022]. }
              \label{fig:typeIa}%
    \end{figure*}

\section{Results}
\label{sec:results}
\subsection{Scan over SNR parameter space}
\label{sec:scan}

As the properties of the molecular clouds are known (table \ref{tab:clouds}), we vary the properties of a hypothetical SNR to investigate the required values to account for the $\gamma$-ray flux of LHAASO\,J2108+5157. We assume that the SNR is located at the same distance from Earth as the cloud. 
The SNR age is varied in ten logarithmically spaced steps between 1\,kyr and 500\,kyr, for a fixed separation distance between the SNR and cloud of 24\,pc. Similarly, the separation distance is independently varied in ten logarithmically spaced steps between 10\,pc and 500\,pc for a fixed SNR age of 4\,kyr. For type Ia supernovae, the fixed reference values were reduced to 10\,pc and 1\,kyr.
These curves are shown in figures \ref{fig:typeII} and \ref{fig:typeIa} for type II and type Ia supernovae respectively. 

In the case of type II supernova remnants shown in figure \ref{fig:typeII}, the predicted flux is comparable to the data for cloud FKT[2022], yet the flux predicted for MML[2017]4607 consistently falls below the measured flux. Younger ages are preferred, with the flux at energies $\lesssim 1$\,TeV becoming over predicted between $\sim$7\,kyr and 30\,kyr for FKT[2022].
The key features of the model are that the highest energy particles escape the SNR at earlier times and are first to arrive at the cloud. The spectral energy distribution hence rises at the highest energies at earlier times (and for shorter distances). The particle distribution then cools as a function of age, with the peak shifting towards lower energies. 

In the case of type Ia supernova remnants shown in figure \ref{fig:typeIa}, a separation distance larger than 24\,pc is required for FKT[2022] to avoid over predicting the flux in the $<10$\,TeV range. MKT[2022]4607 is better able to account for the gamma-ray flux under the type Ia scenario, yet only for an optimum combination of low distance and young age. 
As $t_{\rm sed}$ is lower for the type Ia scenario, the spectral energy distribution is more highly populated at an earlier stage. 

\subsection{Contribution from the galactic CR sea}
\label{sec:CRsea}
As described above, the contribution from diffusive galactic CRs is included in the model, assuming that the particle flux is comparable to that measured at Earth \citep{PhysRevLett.114.171103_AMS}. From the parameter scan, we find that the contribution from the nearby SNR dominates over that from galactic CR sea in most cases. Indeed, the contribution from diffusive galactic CRs only exceeds that from the SNR if either the cloud-SNR distance is $\gtrsim200$\,pc (for young $\lesssim 10$\,kyr SNRs), or if the SNR is old, $\gtrsim 400$\,kyr (for nearby $\lesssim 50$\,pc SNRs). 

In order to test whether the diffuse galactic CR sea could be solely responsible for the measured gamma-ray flux, the normalisation of the galactic flux contribution was varied, in the absence of considering any hypothetical SNR. To match the observed emission at TeV energies using the molecular clouds considered, the normalisation must be of order $\sim 10^3$ higher than that measured at Earth. This enhancement is unlikely to be achieved without the presence of an accelerator nearby. 

Next, we consider all possible combinations of SNR age and separation distance within the aforementioned ranges. A chisquare evaluation of the model curve to the LHAASO data points only is used to establish which model curves provide the closest match to the data. Due to the large number of free parameters entering into the model, we do not perform a minimisation, as there will be multiple local minima in the parameter space able to account for the data. Rather, we aim provide a plausible range of allowed values for the specific case of this model, with assumed fixed parameters as in table \ref{tab:modelpars}. 

   \begin{figure*}
   \centering
   \includegraphics[width=\columnwidth]{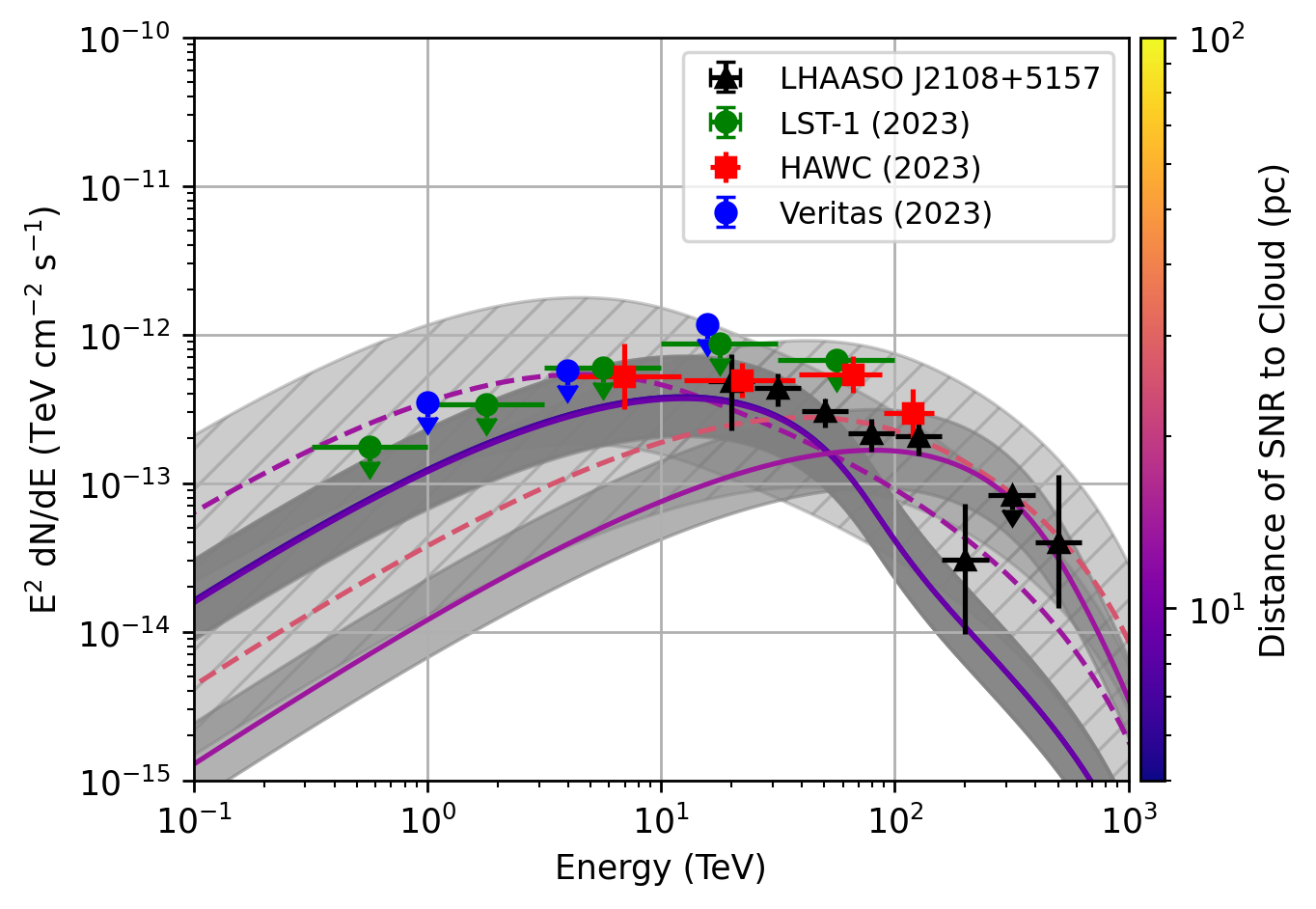}
   \includegraphics[width=\columnwidth]{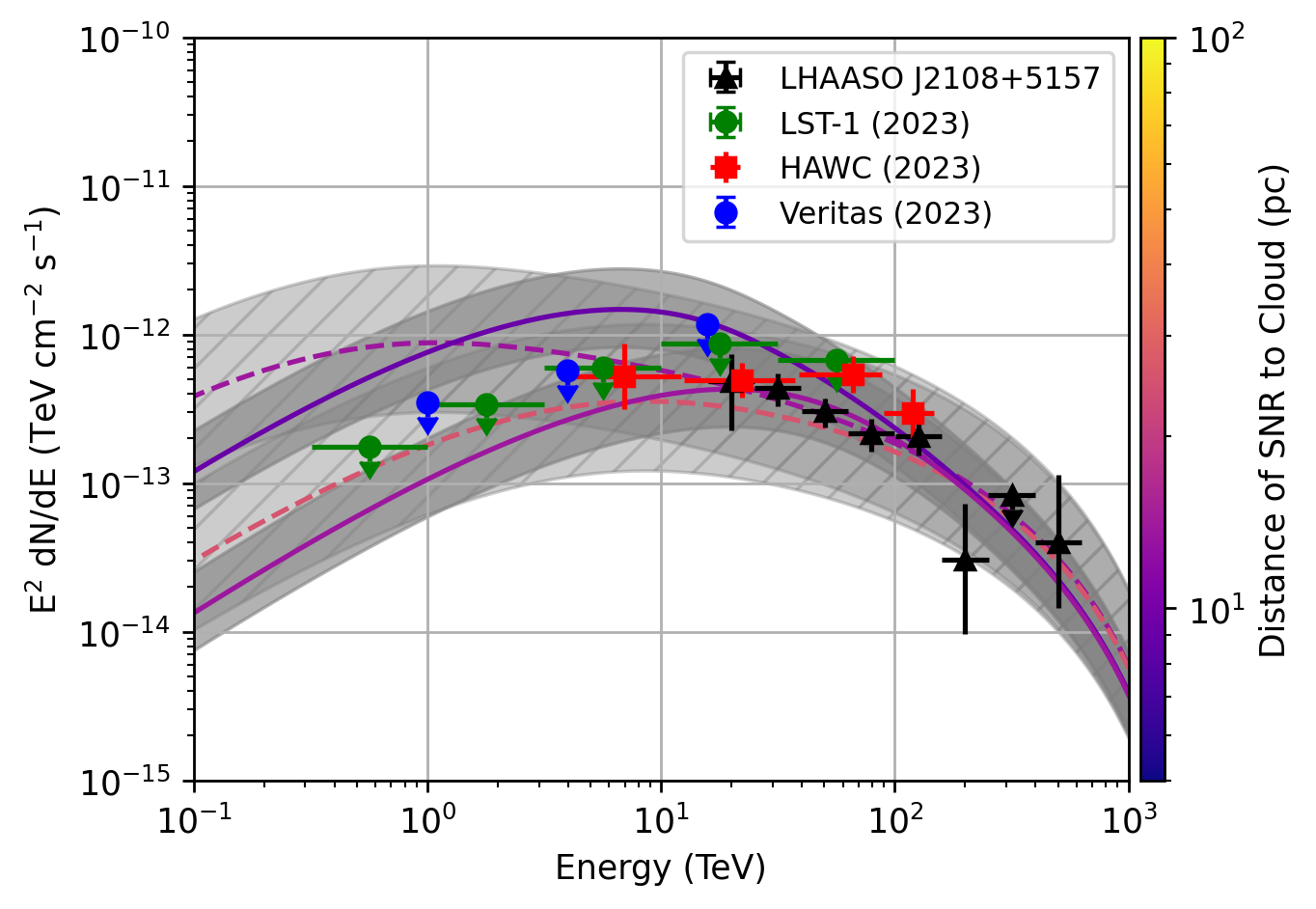}
      \caption{Model curves corresponding to the parameter combinations that best match the data as listed in table \ref{tab:best}. Left: type II and Right: type Ia supernova remnant. Solid lines and shaded uncertainty band correspond to MML[2017]4607. Dashed lines and hatched region correspond to FKT[2022].}
    \label{fig:typeIIbest}
   \end{figure*}
%-----------------------------------------------------------------

\subsection{Best-matched models}
\label{sec:best}
\subsubsection{Clouds MML[2017]4607 and FKT[2022]}

For each cloud, model curves corresponding to the two best matching combinations of SNR age and separation distance are shown in figure \ref{fig:typeIIbest}. Model curves for MML[2017]4607 were consistently below the data points for $\chi=0.1$ within the cloud, as seen in figures \ref{fig:typeII} and \ref{fig:typeIa}. To obtain parameter values within comparable agreement to the data as for FKT[2022], we neglected the suppressed diffusion within the cloud for MML[2017]4607 (and for this section only) by setting $\chi=1$. This corresponds to the most optimistic case in which CRs can freely penetrate the cloud, although we note that $\delta$ was kept fixed to $0.5$ and we did not investigate the effect of altering the energy dependence of the diffusion coefficient in equation \eqref{eq:diffcoeff}. 

The best matching combinations are summarised in table \ref{tab:best}. 
In general, the SNR age was found to have a stronger influence on the curve shape and hence quality of the match to LHAASO data than the separation distance. 
FKT[2022] yielded more parameter combinations with a lower $\chi^2$ than MML[2017]4607, where model curves for the same age yet for smaller distances were essentially consistent. 
This is supported by figures \ref{fig:typeII} and \ref{fig:typeIa} - for a fixed age, provided the distance is sufficiently small that CRs have had time to traverse the cloud, the gamma-ray flux remains constant with decreasing distance. (Equivalently, the gamma-ray flux drops with increasing distance.) 
Overall, the type Ia scenario (i.e. a lower $t_{\rm sed}$) is preferred.

One might ask whether or not a finer-resolution of values covering the reasonable parameter space would lead to a model that better matched the data. Whilst this may be the case, we first consider the effect of propagating the uncertainties in the measured properties of the clouds (table \ref{tab:clouds}) through the model. An upper bound to the flux is obtained by adopting the 1\,$\sigma$ deviation $d-\sigma_d$ and $n+\sigma_n$, whilst a lower bound is similarly obtained from the model evaluated with $d+\sigma_d$ and $n-\sigma_n$, where we intrinsically assume that the uncertainties are Gaussian distributed. Increasing $n$ will increase the target material and hence flux as per equation \eqref{eq:gflux}, whilst increasing $d$ will decrease the flux as per equation \eqref{eq:cloudflux}. 
For FKT[2022] uncertainties are reported in \cite{2023PASJ...75..546DeLaFuente}, whilst for MML[2017]4607 uncertainties are not provided in the case that near and far estimates agree \citep{MDsurvey17}. We therefore adopt a 20\% uncertainty in $d$ and $n$ for MML[2017]4607 as a rough estimate, given that the true uncertainty and subsequent variation in the model is unknown.  
Resulting uncertainty bands corresponding to the parameter combinations reported in table \ref{tab:best} are shown in figure \ref{fig:typeIIbest}.

    \begin{threeparttable}
        \renewcommand{\TPTminimum}{\columnwidth}
       \caption{Combinations of SNR age, $t$ and separation distance, $\Delta d$ for the model curves that best match the LHAASO data, listed in ranked order. These curves are shown in figure \ref{fig:typeIIbest}. }
    \begin{tabular}{lcccc}
        Cloud & $t$ (kyr) & $\Delta d$ (pc) & SN type & $\chi^2$ \\
        \hline
        MML[2017]4607 & 1 & 37 & Ia & 5.1 \\
        FKT[2022] & 4 & 37 \tnote{*} & Ia & 6.7 \\
        FKT[2022] & 4 & 57 & Ia & 9.2 \\
        FKT[2022] & 4 & 57 & II & 15.5 \\
        FKT[2022] & 8 & 24 \tnote{**} & II & 17.0 \\
        MML[2017]4607 & 4 & 24 \tnote{**} & II & 24.4 \\
        MML[2017]4607 & 2 & 37 & II & 25.0 \\
        MML[2017]4607 & 1 & 24 & Ia & 28.2 \\
        \hline
    \end{tabular}
    \begin{tablenotes}
    \small
    \item[*] Model curves for the same SNR age yet with smaller distances provided a comparable fit to the LHAASO data, but severely overestimated the LST-1 upper limits and are hence not shown. 
    \item[**] Model curves for the same SNR age yet with a distance of 10\,pc and 15\,pc were comparable to the 24\,pc distance quoted.
    \end{tablenotes}
    \vspace{5mm}
    \label{tab:best}
    \end{threeparttable}

%----------------------------------------------------------------- 

Figure \ref{fig:typeIIbest} clearly illustrates that the uncertainty introduced to the model from experimental measurements (or the adopted 20\% uncertainty) on the cloud properties leads to variation in predicted flux comparable to that seen by varying the input age and distance of the parameter scan. Therefore, a more finely-grained exploration of the SNR parameter space is not well-motivated.

\subsubsection{Corresponding neutrino flux}

\begin{figure}
   \centering
   \includegraphics[width=\columnwidth]{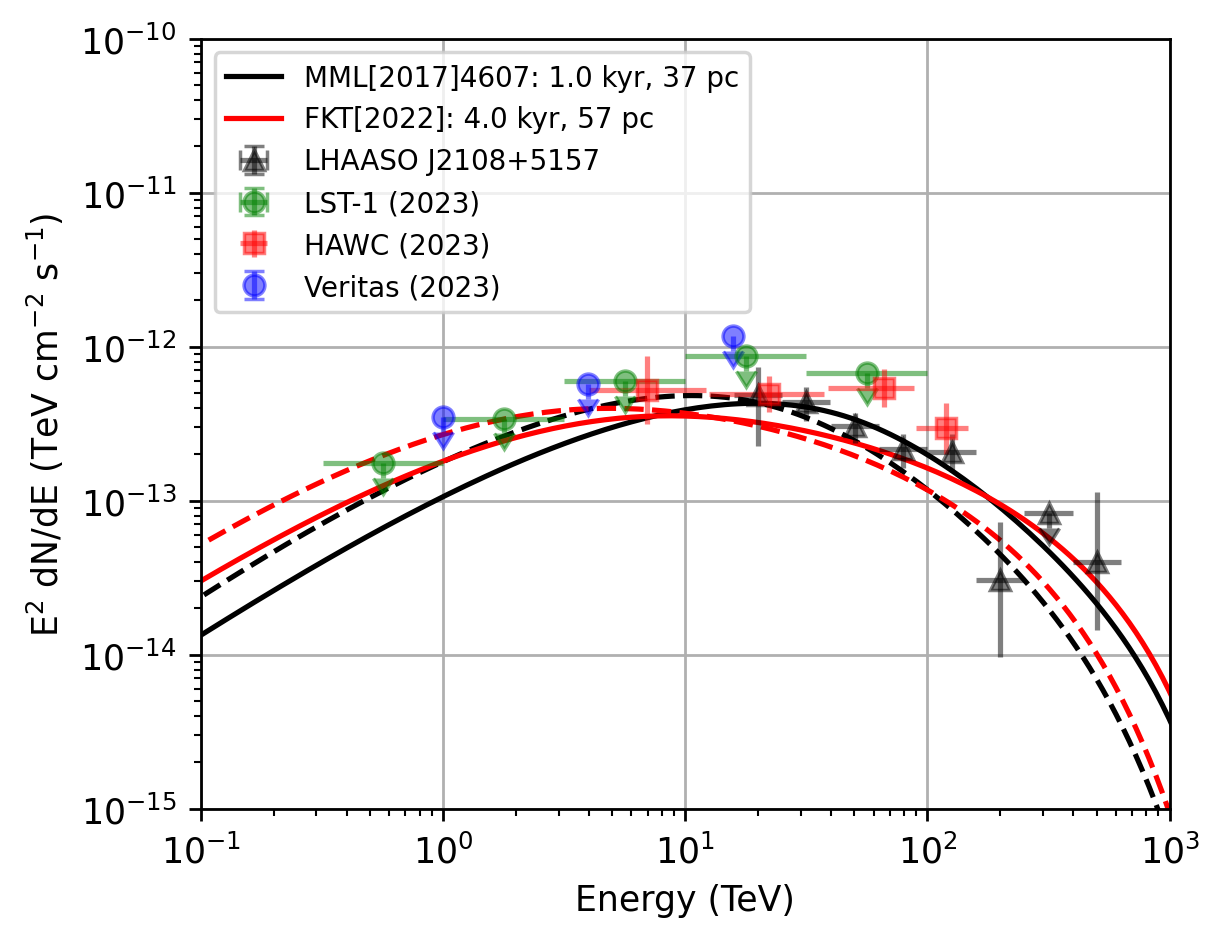} %3pev
      \caption{ Gamma-ray and neutrino fluxes for two of the best-matching scenarios for a type Ia SN from table \ref{tab:best}. Solid lines correspond to the expected gamma-ray flux and dashed lines to the neutrino flux. }
         \label{fig:typeIInu}
   \end{figure}

For two of the best matching models from table \ref{tab:best}, we show the corresponding total neutrino flux in figure \ref{fig:typeIInu}. For MML[2017]4607 this is for $t=1$\,kyr and $\Delta d=37$\,pc, whilst for FKT[2022] we show $t=4$\,kyr and $\Delta d=57$\,pc, both for the SN Ia case. Although $\Delta d=37$\,pc yielded a lower $\chi^2$ for FKT[2022] with respect to the LHAASO data, this curve is disfavoured as it exceeds the upper limits provided by LST-1 (upper dashed curve in figure \ref{fig:typeIIbest}).
These curves essentially scale with the $\gamma$-ray flux, yet still lie at least an order of magnitude in flux below the sensitivity of current neutrino experiments suited for the detection of astrophysical neutrinos, such as IceCube \citep{2019ApJ...886...12A_IceCube}. 

\section{Discussion}
\label{sec:discuss}

LHAASO\,J2108+5157 is an intriguing UHE gamma-ray source with no known counterparts yet spatially coincident with molecular clouds. In this study, we investigate a scenario whereby the molecular cloud is illuminated by energetic protons accelerated at a SNR in the vicinity. By scanning the parameter space of SNR age and separation distance between the hypothetical SNR and the cloud, we obtain model predictions that can be compared to data, thereby constraining the most likely SNR properties. Consistently, we find that a comparatively young ($<10$\,kyr) and nearby ($d\lesssim40-60$\,pc) SNR is required. 

There are currently no known SNRs matching this description. From the SNR catalogue \citep{snrcat}, the two closest SNRs are G094.0+01.0 and G093.7-00.2, at angular distances of more than $2.4^\circ$ from LHAASO\,J2108+5157. At the 3.28\,kpc distance of MML[2017]4607 this corresponds to 140\,pc and 190\,pc separation from the cloud respectively, whilst at the 1.7\,kpc distance of FKT[2022] the SNRs are situated 80\,pc and 110\,pc away from the cloud. Additionally, G094.0+1.0 has an estimated age of 25\,kyr, far older than the SNR ages preferred by our model. We conclude that neither SNR is associated to LHAASO\,J2108+5157. 

Nevertheless, it remains plausible that there are further, as yet undiscovered SNRs located in the region. Recent results from the EMU/POSSUM survey, performed using the Australian Square Kilometer Array Pathfinder (ASKAP) observed a region of the galactic plane containing 7 known SNRs, and found 21 candidates, of which 13 were new discoveries \cite{2023MNRAS.524.1396Ball_askap}. This supports the notion that radio surveys to date may not be sufficiently sensitive to detect all SNRs within a given region.

Several molecular clouds have been identified in the region, two from \cite{MDsurvey17} based on \cite{Dame01} (MML[2017]4607 and MML[2017]2870) and most recently a new cloud FKT[2022] reported by \cite{2023PASJ...75..546DeLaFuente}. Model parameters were explored for the two clouds spatially coincident with LHAASO\,J2108+5157, namely MML[2017]4607 and FKT[2022].

Both type II and type Ia supernova explosion scenarios were considered, where the main difference is in the assumed time for transition to the Sedov-Taylor phase ($t_{\rm sed}$). 
Although a better match could be achieved under the type Ia scenario, we consider this unlikely. Type Ia supernoave occur in older systems where at least one member of a binary system has sufficiently evolved to become a white dwarf, generally corresponding to environments not rich in molecular material. Type II supernovae, however, occur in younger environments where an abundance of molecular material can be expected, similar to that observed in the vicinity of LHAASO\,J2108+5157. Hence, we rather interpret these results as indicating that an earlier transition into the Sedov-Taylor phase is preferred, which may reflect (e.g.) properties of the ambient medium rather than the nature of the progenitor \citep{1999ApJS..120..299Truelove}. 

In all model curves, the highest energy data point at $\sim$500\,TeV could not be well matched with a maximum energy of the proton spectrum of 1\,PeV. Therefore, throughout this study we assumed a maximum energy at the Sedov time of 3\,PeV. 

For MML[2017]4607 to account for the data, we neglected an additional suppression within the cloud due to turbulence compared to the ISM (i.e. $\chi=1$). With $\chi=0.1$ within the cloud, MML[2017]4607 consistently under predicted the data in our model (figures \ref{fig:typeII} and \ref{fig:typeIa}). Our model assumed locally suppressed diffusion compared to the Galactic average also in the intervening medium between the SNR and the cloud, a reasonable assumption for regions of active particle acceleration \citep{Gabici07,dangelo,Inoue2019ApJ...872...46I}. 

Suppressed diffusion and young SNR age as preferred model parameters is in agreement with the 4.5\,kyr age obtained by \cite{2022ApJ...926..110Kar}, although \cite{2023MNRAS.521L...5DeSarkar} suggests an older SNR age of 44\,kyr, obtained with a different spectral index for the particle population. A young SNR may still be a comparatively weak producer of synchrotron emission, or could be of small size and remain embedded within (or obscured by) molecular clouds in the region. 
Given the angular size of the molecular clouds, a young SNR could be completely hidden behind the clouds along the line of sight. Using the relation $R_{\rm SNR}\propto t^{2/5}$ for evolution in the Sedov-Taylor phase, an SNR younger than 12\,kyr for MML[2017]2870 and 19\,kyr for FKT[2022] would be small enough to be obscured by the cloud. This is consistent with the preferred $<10$\,kyr SNR age.

Nonetheless, other scenarios for the origin of LHAASO\,J2108+5157 remain plausible. Young stellar clusters have been hypothesised as suitable Galactic PeVatrons, with particle acceleration occurring at the termination shock of the collective wind \citep{Aharonian2019_natast,2021MNRAS.504.6096Morlino}. There are two known young stellar clusters nearby to LHAASO\,J2108+5157: although the distance to Kronberger\,80 is known to be at least 4.8\,kpc or larger \citep{2016A&A...585A.101Karchenko,2020A&A...633A..99Cantat}, disfavouring an association with the molecular clouds in the region, and the distance to Kronberger\,82 remains unknown \citep{2006A&A...447..921Kronberger}. As such, a stellar cluster is a potential alternative accelerator, also capable of illuminating molecular clouds with CRs, but not well motivated in this region. 

Given the spatial correlation of LHAASO\,J2108+5157 with molecular clouds a leptonic scenario for the emission seems unlikely, nevertheless it has been demonstrated that powerful pulsar wind nebulae are capable of accelerating leptons to beyond 1\,PeV and can account for UHE gamma rays, especially in high radiation field environments \citep{2009A&A...497...17Vannoni,2022A&A...660A...8Breuhaus}. The lack of a pulsar counterpart, or of X-ray synchrotron emission that would indicate the presence of a pulsar wind nebula also in cases where the pulsed emission is mis-aligned, disfavours such a scenario. 

With the advent of current generation detectors such as LHAASO sensitive to UHE gamma rays, we may expect other enigmatic sources to emerge, corresponding to clouds illuminated by unknown accelerators. Other unidentified gamma-ray sources for which no known counterpart has been identified to date, such as LHAASO\,J0341+5258, may have a similar origin \citep{2021ApJ...917L...4CaoJ0341}. The first LHAASO catalogue reported no fewer than seven further new sources that seem to be ``dark'' in nature, without any known counterparts \citep{2023arXiv230517030Cao_1lhaaso}. Undoubtedly, further follow-up studies, both in terms of observation and interpretation, are necessary to determine the origin of these enigmatic gamma-ray sources.

\section{Conclusion}
\label{sec:conclusion}
LHAASO\,J2108+5157 is a dark UHE gamma-ray source spatially coincident with two molecular clouds. We find that the gamma-ray emission can be accounted in terms of molecular cloud illumination by CRs from a nearby ($\lesssim40-60$\,pc) young ($<10$\,kyr) SNR. Although no SNR is currently known matching these criteria, such an SNR could be obscured by other material along the line of sight, or simply lie below the detection threshold of previous surveys \citep{2023MNRAS.524.1396Ball_askap}. Interactions of the diffuse galactic CR sea with the molecular clouds is found to be insufficient to explain the observed gamma-ray flux.

As the exposure of current survey instruments increases, and with the advent of future facilities such as CTA and SWGO, we can anticipate further such discoveries, potentially unveiling a population of UHE sources tracing the presence of PeV particles \citep{2013APh....43....3A_CTAintro,2019_SWGO_whitepaper,2022Univ....8..505CasanovaPeV,2023arXiv230517030Cao_1lhaaso}. 
The key to identifying PeVatrons may lie not in emission from the accelerators themselves, but rather from evidence of energetic particles that have escaped the source region. 

\begin{acknowledgements}
The author is grateful to G. Rowell \& C. van Eldik for fruitful discussions and especially to A. Specovius for reading the manuscript.  
This work is supported by the \emph{Deut\-sche For\-schungs\-ge\-mein\-schaft, DFG\/} project number 452934793.
\end{acknowledgements}

% use BibTeX with the regular commands:
\bibliographystyle{aa} % style aa.bst
\bibliography{references} % your references Yourfile.bib

\end{document}